\documentclass[aps,prd,twocolumn,superscriptaddress,showpacs]{revtex4}

\usepackage{epsfig}
\usepackage{graphicx}
\usepackage{dcolumn}
\usepackage{bm}
\usepackage{setspace}

\begin{document}

\title{Measurement of 
$\sigma_{\chi_{c2}} {\cal B}(\chi_{c2} \rightarrow J/\psi \, \gamma)/
 \sigma_{\chi_{c1}} {\cal B}(\chi_{c1} \rightarrow J/\psi \, \gamma)$ in
$p\overline{p}$ Collisions at $\sqrt{s} =$ 1.96 TeV }
\affiliation{Institute of Physics, Academia Sinica, Taipei, Taiwan 11529, Republic of China} 
\affiliation{Argonne National Laboratory, Argonne, Illinois 60439} 
\affiliation{Institut de Fisica d'Altes Energies, Universitat Autonoma de Barcelona, E-08193, Bellaterra (Barcelona), Spain} 
\affiliation{Baylor University, Waco, Texas  76798} 
\affiliation{Istituto Nazionale di Fisica Nucleare, University of Bologna, I-40127 Bologna, Italy} 
\affiliation{Brandeis University, Waltham, Massachusetts 02254} 
\affiliation{University of California, Davis, Davis, California  95616} 
\affiliation{University of California, Los Angeles, Los Angeles, California  90024} 
\affiliation{University of California, San Diego, La Jolla, California  92093} 
\affiliation{University of California, Santa Barbara, Santa Barbara, California 93106} 
\affiliation{Instituto de Fisica de Cantabria, CSIC-University of Cantabria, 39005 Santander, Spain} 
\affiliation{Carnegie Mellon University, Pittsburgh, PA  15213} 
\affiliation{Enrico Fermi Institute, University of Chicago, Chicago, Illinois 60637} 
\affiliation{Comenius University, 842 48 Bratislava, Slovakia; Institute of Experimental Physics, 040 01 Kosice, Slovakia} 
\affiliation{Joint Institute for Nuclear Research, RU-141980 Dubna, Russia} 
\affiliation{Duke University, Durham, North Carolina  27708} 
\affiliation{Fermi National Accelerator Laboratory, Batavia, Illinois 60510} 
\affiliation{University of Florida, Gainesville, Florida  32611} 
\affiliation{Laboratori Nazionali di Frascati, Istituto Nazionale di Fisica Nucleare, I-00044 Frascati, Italy} 
\affiliation{University of Geneva, CH-1211 Geneva 4, Switzerland} 
\affiliation{Glasgow University, Glasgow G12 8QQ, United Kingdom} 
\affiliation{Harvard University, Cambridge, Massachusetts 02138} 
\affiliation{Division of High Energy Physics, Department of Physics, University of Helsinki and Helsinki Institute of Physics, FIN-00014, Helsinki, Finland} 
\affiliation{University of Illinois, Urbana, Illinois 61801} 
\affiliation{The Johns Hopkins University, Baltimore, Maryland 21218} 
\affiliation{Institut f\"{u}r Experimentelle Kernphysik, Universit\"{a}t Karlsruhe, 76128 Karlsruhe, Germany} 
\affiliation{High Energy Accelerator Research Organization (KEK), Tsukuba, Ibaraki 305, Japan} 
\affiliation{Center for High Energy Physics: Kyungpook National University, Taegu 702-701, Korea; Seoul National University, Seoul 151-742, Korea; and SungKyunKwan University, Suwon 440-746, Korea} 
\affiliation{Ernest Orlando Lawrence Berkeley National Laboratory, Berkeley, California 94720} 
\affiliation{University of Liverpool, Liverpool L69 7ZE, United Kingdom} 
\affiliation{University College London, London WC1E 6BT, United Kingdom} 
\affiliation{Centro de Investigaciones Energeticas Medioambientales y Tecnologicas, E-28040 Madrid, Spain} 
\affiliation{Massachusetts Institute of Technology, Cambridge, Massachusetts  02139} 
\affiliation{Institute of Particle Physics: McGill University, Montr\'{e}al, Canada H3A~2T8; and University of Toronto, Toronto, Canada M5S~1A7} 
\affiliation{University of Michigan, Ann Arbor, Michigan 48109} 
\affiliation{Michigan State University, East Lansing, Michigan  48824} 
\affiliation{Institution for Theoretical and Experimental Physics, ITEP, Moscow 117259, Russia} 
\affiliation{University of New Mexico, Albuquerque, New Mexico 87131} 
\affiliation{Northwestern University, Evanston, Illinois  60208} 
\affiliation{The Ohio State University, Columbus, Ohio  43210} 
\affiliation{Okayama University, Okayama 700-8530, Japan} 
\affiliation{Osaka City University, Osaka 588, Japan} 
\affiliation{University of Oxford, Oxford OX1 3RH, United Kingdom} 
\affiliation{University of Padova, Istituto Nazionale di Fisica Nucleare, Sezione di Padova-Trento, I-35131 Padova, Italy} 
\affiliation{LPNHE, Universite Pierre et Marie Curie/IN2P3-CNRS, UMR7585, Paris, F-75252 France} 
\affiliation{University of Pennsylvania, Philadelphia, Pennsylvania 19104} 
\affiliation{Istituto Nazionale di Fisica Nucleare Pisa, Universities of Pisa, Siena and Scuola Normale Superiore, I-56127 Pisa, Italy} 
\affiliation{University of Pittsburgh, Pittsburgh, Pennsylvania 15260} 
\affiliation{Purdue University, West Lafayette, Indiana 47907} 
\affiliation{University of Rochester, Rochester, New York 14627} 
\affiliation{The Rockefeller University, New York, New York 10021} 
\affiliation{Istituto Nazionale di Fisica Nucleare, Sezione di Roma 1, University of Rome ``La Sapienza," I-00185 Roma, Italy} 
\affiliation{Rutgers University, Piscataway, New Jersey 08855} 
\affiliation{Texas A\&M University, College Station, Texas 77843} 
\affiliation{Istituto Nazionale di Fisica Nucleare, University of Trieste/\ Udine, Italy} 
\affiliation{University of Tsukuba, Tsukuba, Ibaraki 305, Japan} 
\affiliation{Tufts University, Medford, Massachusetts 02155} 
\affiliation{Waseda University, Tokyo 169, Japan} 
\affiliation{Wayne State University, Detroit, Michigan  48201} 
\affiliation{University of Wisconsin, Madison, Wisconsin 53706} 
\affiliation{Yale University, New Haven, Connecticut 06520} 
\author{A.~Abulencia}
\affiliation{University of Illinois, Urbana, Illinois 61801}
\author{J.~Adelman}
\affiliation{Enrico Fermi Institute, University of Chicago, Chicago, Illinois 60637}
\author{T.~Affolder}
\affiliation{University of California, Santa Barbara, Santa Barbara, California 93106}
\author{T.~Akimoto}
\affiliation{University of Tsukuba, Tsukuba, Ibaraki 305, Japan}
\author{M.G.~Albrow}
\affiliation{Fermi National Accelerator Laboratory, Batavia, Illinois 60510}
\author{D.~Ambrose}
\affiliation{Fermi National Accelerator Laboratory, Batavia, Illinois 60510}
\author{S.~Amerio}
\affiliation{University of Padova, Istituto Nazionale di Fisica Nucleare, Sezione di Padova-Trento, I-35131 Padova, Italy}
\author{D.~Amidei}
\affiliation{University of Michigan, Ann Arbor, Michigan 48109}
\author{A.~Anastassov}
\affiliation{Rutgers University, Piscataway, New Jersey 08855}
\author{K.~Anikeev}
\affiliation{Fermi National Accelerator Laboratory, Batavia, Illinois 60510}
% ...
\collaboration{CDF Collaboration\footnote{With visitors from $^a$University of Athens, 
$^b$University of Bristol, 
$^c$University Libre de Bruxelles, 
$^d$Cornell University, 
$^e$University of Cyprus, 
$^f$University of Dublin, 
$^g$University of Edinburgh, 
$^h$University of Heidelberg, 
$^i$Universidad Iberoamericana, 
$^j$University of Manchester, 
$^k$Nagasaki Institute of Applied Science, 
$^l$University de Oviedo, 
$^m$University of London, Queen Mary and Westfield College, 
$^n$Texas Tech University, 
$^o$IFIC(CSIC-Universitat de Valencia), 
}}
\noaffiliation

\pacs{13.60Le, 13.85Qk}

\begin{abstract}
We measure the ratio of cross section times branching fraction,
$R_p \equiv
 \sigma_{\chi_{c2}} {\cal B}(\chi_{c2} \rightarrow J/\psi \, \gamma)/
 \sigma_{\chi_{c1}} {\cal B}(\chi_{c1} \rightarrow J/\psi \, \gamma)$,
in 1.1 fb$^{-1}$ of 
$p\overline{p}$ collisions at $\sqrt{s} =$ 1.96 TeV.
This measurement covers the kinematic range
$p_T( J/\psi)>4.0$ GeV/$c$, $|\eta(J/\psi)| < 1.0$, and 
$p_T(\gamma)>1.0$ GeV/$c$.
For events due to prompt processes, we find 
$R_p = 0.395\pm0.016(stat.)\pm0.015(sys.)$. 
This result represents a significant improvement in precision over previous
measurements of prompt $\chi_{c1,2}$ hadroproduction.
\end{abstract}

\maketitle


Since it was first observed, charmonium ($c\overline{c}$) production
in hadronic collisions has been a subject of considerable
theoretical interest.
Recent approaches to understanding charmonium production make use
of nonrelativistic QCD \cite {Bodwin,Kniehl} to calculate
hadro-production rates at the Tevatron and elsewhere.
While most experimental observations 
of charmonium production consist of $J/\psi$ measurements,
a significant contribution of $J/\psi$ production is indirect,
resulting from the decay of higher mass states \cite {CDF_chi}.
In particular, the radiative decay of the $\chi_{cJ}$ 
states \cite{Footnote_1}
accounts for a significant fraction of the $J/\psi$ production
seen in hadronic collisions, and any calculation of $J/\psi$ production
must include $\chi_{cJ}$ production as well.

Measurements of hadronic $\chi_{cJ}$ production have been made in a 
variety of
beam types and energies \cite {Fixed}
by observing the decay process 
$\chi_{cJ} \rightarrow$ $J/\psi \, \gamma$.  
Experimental results available until now have
suffered from large statistical uncertainties,
and no measurement has had the precision to 
test the consistency of the 
cross-section ratio $\sigma_{\chi_{c2}}/\sigma_{\chi_{c1}}$
with the simple spin-state counting expectation of 
$\frac{5}{3}$ for $\chi_{cJ}$ mesons that are directly
produced in the interaction \cite{Brambilla}.
Knowledge of this ratio is needed in calculations of $J/\psi$ 
production through radiative $\chi_{cJ}$ 
decay, and can be an important standard for comparing production 
models.  

In this letter, we report a measurement of
the relative cross section times branching fractions of the $\chi_{c1}$ and 
$\chi_{c2}$ mesons produced in $p\overline{p}$ collisions 
at a center of mass energy of 1.96 TeV using the
CDF II detector at the Fermilab Tevatron.  We study the inclusive process 
$p\overline{p} \rightarrow \chi_{cJ}X$, where 
$\chi_{cJ} \rightarrow$ $J/\psi \, \gamma$, and
$J/\psi \rightarrow \mu^+\mu^-$, in a data sample with a time-integrated luminosity of 1.1 fb$^{-1}$.  
The final state photon is
reconstructed through its conversion into $e^+e^-$, which provides 
the mass resolution needed to distinguish the $\chi_{c1}$ and $\chi_{c2}$
states. The spatial resolution of the $\mu^+\mu^-$ vertex 
allows separation of prompt $\chi_{cJ}$ production from events 
where the $\chi_{cJ}$ meson is a 
$B$-hadron decay product.
We measure the ratio of the cross section times branching fraction
$R_p \equiv 
     \sigma_{\chi_{c2}}  
     {\cal B}(\chi_{c2}\rightarrow J/ \psi \, \gamma)/
     \sigma_{\chi_{c1}}  
     {\cal B}(\chi_{c1}\rightarrow J/ \psi \, \gamma)$ for promptly
produced $\chi_{cJ}$ mesons.  
In addition, we obtain a result for the analogous
quantity in $B$ decay events, 
$R_B \equiv 
     \sigma_{B}  {\cal B}(B \rightarrow \chi_{c2} \, X)
                 {\cal B}(\chi_{c2}\rightarrow J/ \psi \, \gamma)/
     \sigma_{B}  {\cal B}(B \rightarrow \chi_{c1} \, X) 
                 {\cal B}(\chi_{c1}\rightarrow J/ \psi \, \gamma)$,
which provides a measurement of
${\cal B}(B \rightarrow \chi_{c2} \, X)/
 {\cal B}(B \rightarrow \chi_{c1} \, X)$ for the $B$ hadrons produced in the 
Tevatron environment.

This analysis makes use of the tracking, 
muon identification, and trigger systems.  
The CDF II detector has been described in detail elsewhere 
\cite{CDF_J/psi,CDF_detector}.  
The tracking system consists of a
seven-layer silicon microstrip detector and an open-cell drift chamber (COT)
that operate inside a solenoid with a 1.4 T magnetic field.
 Muon candidates from the decay $J/\psi \rightarrow \mu^+ \mu^-$
 are  identified by  two sets of drift  chambers located
 outside  the   electromagnetic and   hadronic  calorimeters.
The central muon chambers
cover the pseudorapidity region $ |\eta| <  0.6$, 
and  are sensitive to muons having transverse momentum $p_T > 1.4 $ GeV/$c$
\cite {pt_def}.
A second muon system covers the region $  0.6 < |\eta| <  1.0$ and
is sensitive to muons having $p_T > 2.0 $ GeV/$c$.  Muon 
triggering and identification
are based on matching tracks measured in the muon system to COT tracks.

The analysis of the data begins with a selection of well measured
$J/\psi \rightarrow \mu^+ \mu^-$ candidates.
These are selected
 by requiring events that contain two
 oppositely  charged  muon candidates, each with a match between the 
 COT and muon chamber tracks.
 We also require that both muon tracks have 
measurements in at least three layers of the silicon detector and
a two-track invariant  mass within $\pm$80 MeV/$c^2$ of the
world-average  $J/\psi$ mass \cite{PDG}.  
The $J/\psi$ candidates are required to fall within kinematic bounds of
$p_T(J/\psi) >$ 4.0  GeV/$c$ and  $|\eta(J/\psi)| <$  1.0,
which correspond to the approximate limits of our acceptance.
A  simultaneous  mass and vertex constrained fit is performed on 
the muon tracks, where
the  dimuon  mass is   constrained  to the  
world-average  $J/\psi$ mass.  

\begin{figure}[p]
\psfig{figure=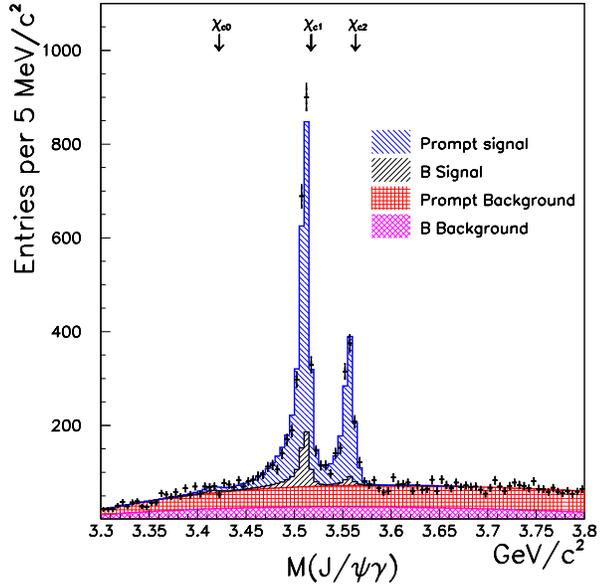, height=3.375in}
\caption{The $J/\psi \, \gamma$ mass distribution (points) with
the projection of the likelihood fit overlaid on the data.
The masses of the $\chi_{cJ}$ mesons and the
contributions of the signal and background components are indicated.
\label{fig:signal} }
\end{figure}

The search for photon conversion 
candidates begins with a scan of all additional tracks 
with $p_T > $ 400 MeV/$c$
found in each $J/\psi$ event.  Two oppositely charged tracks 
are each assigned the electron mass, and 
have their
track parameters recalculated by subjecting them 
and their uncertainties to a fit that has 
constraints consistent with the photon conversion hypothesis.  Specifically,
the two tracks are constrained to be parallel at their point of intersection,
and the momentum vector
of the pair is constrained to originate from the dimuon vertex.
 A   displacement of 12.0 cm or more from the beam line
 in the direction of the
track pair's   transverse  momentum is required
to omit conversions whose momentum is poorly measured due to
bremsstrahlung in the inner detector material.  
We also   require  $p_T(\gamma) >$ 1.0 GeV/$c$.
Finally, a constrained  fit is performed on the four 
tracks that combines the 
$J/\psi$  mass constraint with the photon conversion hypothesis.
The invariant
mass distribution of all $J/\psi \, \gamma$  combinations is shown in
Fig. \ref{fig:signal}, which clearly demonstrates that 
the $J/\psi \, \gamma$
mass resolution acheived by this technique
is sufficient to resolve the  $\chi_{cJ}$ states.

The lifetime of $B$ hadrons allows the transverse
displacement of the dimuon vertex from the beamline to be used
as a tool for their identification.
Since any $J/\psi \, \gamma$ combination that originates from $B$ decay 
represents only a partial reconstruction of the $B$ hadron, the proper
lifetime is not directly measurable.  We therefore use the
quantity 
$ct \equiv 
L \times M(J/\psi)\times F(p_T(J/\psi))/p_T(J/\psi)$
where 
$M(J/\psi$) and $p_T(J/\psi$) are the mass and transverse momentum,
respectively, of the $J/\psi$ candidate, and
$L$ is the measured displacement of the dimuon vertex in the direction
of $p_T(J/\psi$).
The correction factor, $F(p_T(J/\psi))$, was obtained by a Monte Carlo
simulation of $B$ hadron decay \cite{CDF_chi}, 
and provides an average correction between the measured displacement
and the lifetime of the decaying $B$ hadron.

Mass resolution, acceptance, and reconstruction efficiencies for 
$J/\psi \, \gamma$ final states of different invariant mass have been
studied with a Monte Carlo simulation that generates events
uniformly in rapidity and with a transverse momentum
distribution that matches the measured distribution for
$J/\psi$ events \cite {CDF_J/psi}. 
The simulated events were processed through our reconstruction and
analysis algorithms,
and provided templates for the expected signal shape of the
final $J/\psi \, \gamma$ invariant mass distribution as a function
of $p_T(J/\psi)$. 
In particular, the simulated events enabled an estimate of the 
mass resolution and radiative 
tail due to scattering and radiation by the conversion electrons as they
passed through the material in the detector.

\begin{figure}[p]
\psfig{figure=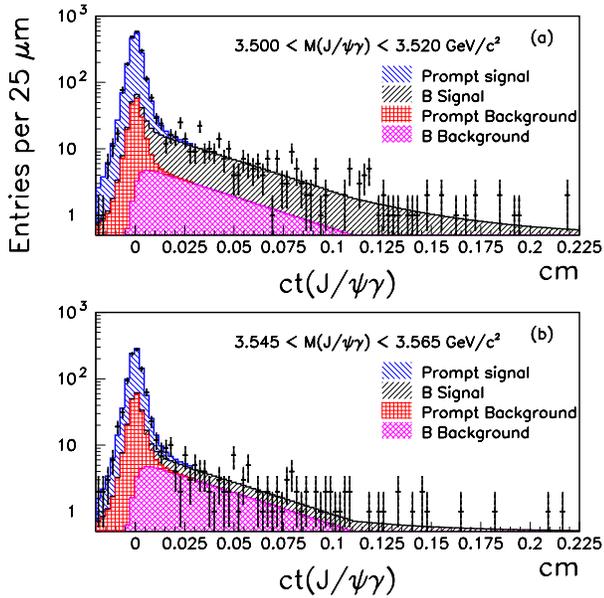, height=3.375in}
\caption{The $ct$  distribution (points) for 
events in the $\chi_{c1}$ (a) and $\chi_{c2}$ (b) 
mass ranges.  
The projection of the fit is overlaid on the data,  
with the contribution of each signal and background component indicated.
\label{fig:life} }
\end{figure}

We used an unbinned likelihood fit to calculate the yield of
$\chi_{cJ}$ events for both prompt and $B$-decay production processes.  
The probability density function used for the fit is a function of both 
$J/\psi \, \gamma$ invariant mass and 
$ct$.  Independent signal and background distributions
are calculated for both processes.
The mass distributions of the signals are constrained to the 
templates obtained 
through simulation.
The mass distributions of the backgrounds
are modeled by polynomials, and 
the probability density function for each event uses
the calculated uncertainty on the invariant mass and $ct$.
The $ct$ distribution is used to separate the production processes,
and is modeled as a sum of prompt (Gaussian resolution) and
$B$-decay (exponential, convoluted with resolution) contributions.
Our fit to the data gives an event yield $(N_{\chi_{cJ}})$ of 
$N_{\chi_{c0}}=41\pm20$,
$N_{\chi_{c1}}=2143\pm60$, and $N_{\chi_{c2}}=1035\pm40$ for promptly produced
events.  For $B$ decay events, the yields are 
$N_{\chi_{c0}}=29\pm16$,
$N_{\chi_{c1}}=384\pm35$, and $N_{\chi_{c2}}=66\pm16$.
Projections of the unbinned likelihood fit are overlaid
onto the mass and $ct$ distributions shown in 
Figs. \ref{fig:signal} and \ref{fig:life}.
The relatively small yield of $\chi_{c0}$ candidates is due to the small
branching fraction into the $J/\psi \, \gamma$ final state \cite{PDG}, 
and is the reason they were not used in the subsequent analysis.

For our acceptance calculation,
we have analyzed our simulated events assuming every photon converted, and
created electron-positron pairs according to the
Bethe-Heitler distribution \cite{Rossi}.
We then counted the number of events that would have been accepted
if all final state products were to pass our kinematic requirements.
Our low transverse momentum limit of 400 MeV/$c$ for the electrons
results in a dependence of our acceptance on the invariant mass of the 
parent.  
The overall ratio of 
the $\chi_{c1}$ and  $\chi_{c2}$ meson acceptances 
($\epsilon_{\chi_{c2}}/\epsilon_{\chi_{c1}}$)  is listed in 
Table \ref{table:final_results} for
several ranges of $p_T(J/\psi)$.  The acceptance ratio is then combined with
our yield ratios to provide measurements of
$R_p$ and $R_B$ for several ranges of $p_T(J/\psi)$.

\begin{table}[hbt]
\begin{center}
\caption{The acceptance ratio and ratios of cross section times 
branching fractions of the $\chi_{cJ}$ states for the prompt events
and $B$ decay events.  
Uncertainties listed are statistical only.
 \protect\label{table:final_results}}
\vspace{3mm}
\begin{tabular}{cccc}
\hline \hline
$p_T(J/\psi$) (GeV/$c$)& $\epsilon_{\chi_{c2}}/\epsilon_{\chi_{c1}}$ &
 $R_p$ & $R_B$ \\
\hline
$4-6$   & $1.27\pm0.01$ & 0.457$\pm$0.039 & 0.150$\pm$0.087 \\
$6-8$   & $1.17\pm0.01$ & 0.384$\pm$0.034 & 0.080$\pm$0.094 \\
$8-10$  & $1.14\pm0.01$ & 0.455$\pm$0.053 & 0.116$\pm$0.070 \\
$ > 10$ & $1.10\pm0.01$ & 0.309$\pm$0.045 & 0.197$\pm$0.082 \\
\hline
$ > 4$  & $1.23\pm0.01$ & 0.395$\pm$0.016 & 0.143$\pm$0.042 \\
\hline \hline
\end{tabular}
\end{center}
\end{table}

Several systematic effects that might change the
reconstruction efficiency ratio 
$\epsilon_{\chi_{c2}}/\epsilon_{\chi_{c1}}$
were studied.
First, the simulated event sample size used for the acceptance
calculation provides an overall relative uncertainty of $\pm0.005$ on the ratio.
A comparison between the full event simulation/reconstruction and
the simpler simulation based on the electron energy distribution
yields a relative systematic uncertainty of $\pm0.020$.
Another effect considered is that
polarization of one of the two $\chi_{cJ}$ states would also introduce
a systematic shift.  We have evaluated the effect of having one state
decay with a distribution given by 
$I(\theta_{\mu\mu}) \propto 1 + \alpha \cos^2(\theta_{\mu\mu})$
where $\theta_{\mu\mu}$ is the polar angle of the $\mu^+$ in the
$J/\psi$ rest frame, and
we take $\alpha=0.13 \pm 0.15$ as was done for a separate
$J/\psi$ cross-section measurement \cite{CDF_J/psi}.  
A variation of $\alpha$ 
by this uncertainty for one of the two $\chi_{cJ}$ states 
implies a relative shift of $\pm0.030$ on our reconstruction efficiency ratio.

We have also considered the sources of systematic uncertainty in the
yield ratio 
calculation.
We varied the invariant mass signal shape 
obtained from the simulation within 
the uncertainty of its parametrization, and found that the relative shift
in the yield ratio is $\pm0.005$.
The uncertainty on parameters used in the $ct$ definition 
corresponds to a variation of 
the $B$ fraction of the $J/\psi \, \gamma$ events of
$\pm0.007$ of its value,
giving a systematic uncertainty of $\pm0.002(\pm0.010$)
in the yield ratio for the prompt($B$ decay) sample.
Finally, we explored the possibility that our data contain partially
reconstructed $h_c \rightarrow J/\psi \, \pi^0$, 
$\pi^0 \rightarrow \gamma \gamma$ events.  
This has been studied by 
simulating this process, parametrizing the resulting $J/\psi \, \gamma$ 
invariant mass distribution, and including this possible background in our
likelihood fits for the signal yield.  The possible
$h_c$ background contribution
was found to be negligible in this data sample, so no systematic uncertainty
was assigned for this process.

Differences in the two states due to production angular or $p_T$ distributions
would require different production mechanisms for the 
$\chi_{c1}$ and $\chi_{c2}$ mesons, and are, therefore, considered to be unlikely.  
Consequently, we did not assign a systematic 
uncertainty on the $\chi_{cJ}$ acceptance due to production dynamics.
A summary of the systematic uncertainties on the cross section 
times branching fraction ratio
is listed in Table \ref{table:systematic}.
The individual uncertainties are combined in quadrature to give the total
systematic uncertainty.

\begin{table}[hbt]
\begin{center}
\caption{Relative systematic uncertainties on $R_p$ and $R_B$.
\protect\label{table:systematic}}
\vspace{3mm}
\begin{tabular}{cc}
\hline \hline
Effect & Uncertainty \\
\hline
Simulation Sample Size & $\pm0.005$ \\
Photon Conversion Simulation & $\pm0.020$ \\
Polarization Effects & $\pm0.030$ \\
Invariant Mass Resolution & $\pm0.005$  \\
Prompt/B Separation & $\pm0.002 (\pm0.010$ for $B$)    \\
\hline
Total & $\pm0.037$ ($\pm0.038$ for $B$) \\
\hline \hline
\end{tabular}
\end{center}
\end{table}

Our final result on the relative rate of production
for promptly produced $\chi_{cJ}$ states is 
$R_p = 0.395\pm0.016(stat.)\pm0.015(sys.)$ 
for $\chi_{cJ}$ with $p_T(J/\psi)> 4$ GeV/$c$ and
$p_T(\gamma)> 1$ GeV/$c$.  
For $\chi_{cJ}$ resulting from $B$ decay over the same kinematic
range we find
$R_B\ = 0.143\pm0.042(stat.) \pm0.005(sys.)$.
These results provide the most precise 
measurement of the $\chi_{cJ}$ production ratio obtained
in any hadronic interactions.  
Conversion of this measurement into the direct cross-section ratio 
$\sigma_{\chi_{c2}}/\sigma_{\chi_{c1}}$ 
requires a knowledge of the branching fractions, 
which are not measured in this experiment,
and a small correction due to
$\psi(2S) \rightarrow \chi_{cJ} \, \gamma$ decays.
Based on the existing measurement of the prompt $\psi(2S)$ cross section
\cite{CDF_psi2S}, and the $\chi_{cJ}$ contribution to the
prompt $J/\psi$ production cross section \cite{CDF_chi},
we estimate that $4.0\pm1.0(5.0\pm1.0)$\% of our prompt
$\chi_{c1}(\chi_{c2})$ sample is due to decay of promptly produced 
$\psi(2S)$ mesons. 

Prior measurements of the prompt cross-section ratio have been severely
limited in their precision due to the statistical uncertainties inherent to
small data samples \cite{Fixed}.  The relative
precision of previous measurements
has typically been approximately 30\% on the cross-section ratio, and
provides weak guidance for production models.  This work, 
combined with the best branching-fraction ratio measurement 
$R_{J/\psi \, \gamma} \equiv 
                      {\cal B}( \chi_{c1} \rightarrow J/\psi \, \gamma)/
                      {\cal B}( \chi_{c2} \rightarrow J/\psi \, \gamma) =
1.91 \pm 0.10$
available \cite{CLEO_BF}, gives
$R_{p} \, R_{J/\psi \, \gamma} = 
0.75\pm0.03(stat.)\pm0.03(sys.)\pm0.04(B.F.)$, 
where the last 
term in the uncertainty is due to the branching-fraction ratio uncertainty.  
This level of precision should serve to inform any 
future developments in the calculation of hadronic charmonium production.

We thank the Fermilab staff and the technical staffs of the participating
 institutions for their vital contributions. This work was supported by 
the U.S. Department of Energy and National Science Foundation; the Italian
 Istituto Nazionale di Fisica Nucleare; the Ministry of Education, Culture,
 Sports, Science and Technology of Japan; the Natural Sciences and Engineering
 Research Council of Canada; the National Science Council of the Republic of
 China; the Swiss National Science Foundation; the A.P. Sloan Foundation; the
 Bundesministerium f\"ur Bildung und Forschung, Germany; the Korean Science and
 Engineering Foundation and the Korean Research Foundation; 
the Particle Physics and Astronomy Research Council and the Royal Society, UK; 
the Institut National de Physique Nucleaire et Physique des Particules/CNRS; 
the Russian
 Foundation for Basic Research; 
the Comisi\'on Interministerial de Ciencia y
 Tecnolog\'{\i}a, Spain; in part by the European Community's Human Potential
 Programme under contract HPRN-CT-2002-00292; and the Academy of Finland.

\begin{thebibliography}{9}
\bibitem{pt_def}The pseudorapidity $\eta$ is defined as
$\eta \equiv -\ln ( \tan( \theta /2) )$, where $\theta$ 
is the angle between the particle
momentum and the proton beam direction.
\end{thebibliography}
\end{document}